\documentclass[aps,pre,floats, twocolumn,showpacs,superscriptaddress]{revtex4}
\usepackage{graphicx,epsfig}
\usepackage{graphics,dcolumn,bm,epic, eepic,fleqn,float}
\usepackage{amssymb,amsmath,amsfonts,multirow,rotate,color}
\usepackage{soul}
\usepackage{wasysym}
\usepackage{enumerate}

\usepackage{dcolumn}
\usepackage{subfigure}

\DeclareGraphicsExtensions{. jpg,. pdf, . mps, .png, .eps, . ps, . EPS}

\definecolor{Myorange}{cmyk}{0,0.42,1,0}

\newcommand{\lay}[1]{^{[#1]}}

\begin{document}

\def\bc{\begin{center}} 
\def\ec{\end{center}}
\newcommand{\avg}[1]{\langle{#1}\rangle}
\newcommand{\Avg}[1]{\left\langle{#1}\right\rangle}

\title{Non-linear growth and condensation in multiplex networks}

\author{V. Nicosia}
\affiliation{School of Mathematical Sciences, Queen Mary University of
  London, Mile End Road, E1 4NS, London (UK)}
\author{G. Bianconi}
\affiliation{School of Mathematical Sciences, Queen Mary University of
  London, Mile End Road, E1 4NS, London (UK)}
\author{V. Latora}
\affiliation{School of Mathematical Sciences, Queen Mary University of
  London, Mile End Road, E1 4NS, London (UK)}
\author{M. Barthelemy}
\affiliation{Institut de Physique Th\'{e}orique, CEA, CNRS-URA 2306, F-91191, 
Gif-sur-Yvette, France}

\begin{abstract}
Different types of interactions coexist and co-evolve to shape the
structure and function of a multiplex network. We propose here a
general class of growth models in which the various layers of a
multiplex network co-evolve through a set of non-linear preferential
attachment rules. We show, both numerically and analytically, that by
tuning the level of non-linearity these models allow to reproduce
either homogeneous or heterogeneous degree distributions, together
with positive or negative degree correlations across layers.  In
particular, we derive the condition for the appearance of a condensed
state in which one node in each layer attracts an extensive fraction
of all the edges.
\end{abstract}

\maketitle

Various complex systems are well described
by multiplex networks of nodes connected through links of distinct
types, which constitute separate yet co-evolving and interdependent
layers~\cite{Cardillo2013_Emergence,Cardillo2013_Airports,battiston,Thurner}. Examples
of multiplex structures can be found in social, technological,
transportation and communication systems, and in general wherever a
certain set of elementary units is bound by different kinds of
relationships~\cite{arenas,Aleva:2013,Bianconi2013}. In these systems,
links of different types are intertwined in non-trivial ways, so that
it is not possible to study each layer separately. In particular, a
node can have different degrees at the various layers, so that a hub
at one layer might not be a hub in another
layer~\cite{Nicosia2014_corr} or, conversely, the hubs might tend to
be the same across different layers~\cite{Nicosia2013,kim}. Also, it
has been shown that the presence of an edge at a certain layer of a
multiplex network is often correlated with the presence of the same
edge on another layer, which corresponds to a significant overlap of
links~\cite{Thurner,Cardillo2013_Emergence,Bianconi2013,Halu2013}. Some
recent studies have focused on dynamical processes on multiplexes,
including
percolation~\cite{Buldyrev:2010,Cellai2013,Baxter2013,Min2013_robustness,Bianconi2014_percolation},
diffusion~\cite{Gomez2013,Sole2013,DeDomenico2013_RW,Halu2013_Pagerank},
spreading~\cite{Saumell2012,Granell2013,Min2013_epidemic,Cozzo2013_epidemic},
traffic \cite{Morris:2012}, cascades \cite{Morris:2013}, and
cooperation~\cite{GomezGardenes2012,Jiang2013,Wang2013}, and a few
recent works have suggested that degree
correlations~\cite{Parshani2010,Lee2012,Valdez2013} as well as overlap
of links~\cite{Cellai2013,Li,Hu} may have a substantial impact on the
emergence and stability of collective behaviors in multiplex systems.

It is therefore interesting to investigate the mechanisms responsible
for the appearance of inter-layer correlations in multiplexes. A few
different approaches for the modelling of multiplex networks have been
recently proposed. Some of them aim at defining appropriate static
null-models for multiplexes~\cite{arenas,Aleva:2013,Bianconi2013},
while some other focus on capturing the non-equilibrium nature of
multiplexes and on providing possible physical explanations for their
formation~\cite{Nicosia2013}. However, until now, all the existing
models for growing multiplexes with homogeneous and heterogeneous
degree distributions allow for positive inter-layer degree
correlations only.

In this Article, we propose and study a general growth model of
multiplex networks based on a non-linear preferential attachment
mechanism. Using both analytical and numerical arguments, we show that
this model generates different regimes and displays a transition
towards a condensed state where only a few hubs dominate the degree
distribution of each layer. Moreover, in the non-condensed regime the
model can generate multiplexes with homogeneous or heterogeneous
degree distributions, having either positive or negative inter-layer
degree correlations. Finally we notice that in the multi-layer version
of non-linear preferential attachment the structure of the network
dramatically depends on fluctuations, and that the mean-field
approach, which was fundamental to understand network growth in
single-layer networks, actually fails to a large extent in predicting
the dynamics of the growth process. 

The paper is organized as follows. In Sec.~\ref{sec:model} we define a
general class of non-linear preferential attachment growth models for
multi-layer networks focusing, as an example, on the case of a 2-layer
multiplex. In this case, the growth is completely determined by the
relative values of two attaching exponents, called $\alpha$ and
$\beta$. In Sec.~\ref{sec:seminonlinear} we investigate the role of
the exponent $\beta$ when one of the two terms of the attaching kernel
is linear, i.e. $\alpha=1.0$. This is a first generalization of the
classical linear preferential attachment model~\cite{ba}. In
Sec.~\ref{sec:meanfield} we derive a mean-field solution for the
proposed class of models, and we show that the mean-field
approximation fails to account for most of the observed structural
properties of the multiplex, in particular regarding the possibility
to obtain negative inter-layer degree correlations. In
Sec.~\ref{sec:master_eq} we present the master equation of the model
and we solve it to derive the conditions for the appearance of a
condensed state. In Sec.~\ref{sec:numerical} we show and discuss the
full phase diagram of the model based on numerical simulations, which
is in perfect agreement with the analytical predictions obtained by
solving the master equation. In Sec.~\ref{sec:cartography} and in
Sec.~\ref{sec:mixed_corr} we discuss, respectively, the effect of the
parameters on the role played by hubs, by means of the recently
introduced multiplex cartography~\cite{battiston}, and the appearance
of mixed degree correlation patterns.  In Sec.~\ref{sec:distance} we
focus on the values of characteristic path length and multiplex
interdependence obtained as a function of $\alpha$ and $\beta$, while
in Sec.~\ref{sec:calibration} we show how the model can be calibrated
in order to reproduce some of the structural properties of two
real-world multiplex networks. In Sec.~\ref{sec:general} we discuss
three possible generalizations of the model to the case of $M$-layer
multiplex networks, providing also the analytical solution for the
boundary of the condensed phase. Finally, in Sec~\ref{sec:conclusions}
we draw our conclusions and we discuss possible future directions of
research in the field of multiplex network modelling.

\section{Model}
\label{sec:model}

Let us consider a multiplex network consisting of $M$ layers, one for
each type of relationship among nodes, defined by the vector of
adjacency matrices $\left\{A\lay{1}, A\lay{2}, \ldots,
A\lay{M}\right\}$, where $A\lay{\ell}=\{a\lay{\ell}_{ij}\}$
and $a\lay{\ell}_{ij}=1$ if and only if node $i$ and node $j$ are
connected by an edge on layer $\ell$. A node $i$ of the network is
characterized by the vector $\bm{k}_i=\{k\lay{1}_i, k\lay{2}_i,
\ldots, k\lay{M}_i\}$ of the degrees of its replicas at each layer,
where $k\lay{\ell}_i=\sum_{j}a\lay{\ell}_{ij}$. We are interested in
the mechanisms which might be responsible for the growth of the
multiplex.  We start from a connected graph with $m_0$ nodes and we
assume that, at each time $t$, a new node $i$ arrives in the graph,
carrying $m\le m_0$ new links in each layer, and that the probability
$\Pi\lay{\ell}_{i\rightarrow j}$ for node $i$ to attach on layer
$\ell$ to an existing node $j$ is a function $f\lay{\ell}$ of the
degrees of $j$ at all layers:
\begin{equation}
  \Pi\lay{\ell}_{i\rightarrow j} \propto
  f\lay{\ell}\left({k\lay{1}_j}, {k\lay{2}_j},
  \ldots, {k\lay{M}_j}\right)
\end{equation}
For the sake of clarity, and without loss of generality, we focus on a
multiplex network with two layers, where we denote by $k_j$ the degree
of node $j$ in layer $1$, and by $q_j$ the degree of $j$ in layer $2$,
and we assume that
\begin{equation}
  \Pi^{[1]}_{i \rightarrow j} \propto f(k_j,q_j), ~~~~\text{and} ~~~~
  \Pi^{[2]}_{i \rightarrow j} \propto f(q_j,k_j).
\label{pi}
\end{equation}
In the context of single-layer networks, non-linear attachment kernels
of the form $f(k_j)=k_j^\alpha$, with $\alpha \ge 0$, have been
introduced in Ref.~\cite{krapivsky}, as a generalization of linear
preferential attachment models~\cite{ba}. We extend this idea to
networks with multiple layers, also allowing for negative exponents to
mimic the case in which new nodes prefer to avoid linking to
high-degree nodes. We adopt the general expression
\begin{equation}
  f(k_j,q_j) =  {k_j^{\alpha}}{q_j^{\beta}}
\label{f}
\end{equation}
where, by tuning the two exponents $\alpha, \beta \in \mathbb{R}$, we
can model different attachment strategies. If the exponents $\alpha$
and $\beta$ in Eq.~(\ref{f}) are both positive (negative), then new
nodes will preferentially link to nodes which are well-connected
(poorly connected) on both layers. Conversely, if $\alpha>0$ and
$\beta<0$ (resp. $\alpha<0$ and $\beta>0$), a new node will be
preferentially linked, in layer 1, with nodes which are well-connected
(resp. poorly-connected) in layer 1 and poorly connected
(resp. well-connected) in layer 2. A specular interpretation holds for
the attachment probability $f(q_j, k_j)$ on layer 2. As we will show
in the following, the attachment probabilities in Eqs.~(\ref{pi})
and~(\ref{f}) are general enough to produce multiplex networks with
different degree distributions, and with positive and negative
correlations between the degrees of a node at the two layers. There
are several possible ways to generalize this model to the case of more
than two layers, and some of them are discussed in
Sec.~\ref{sec:general}.

\section{Semi-nonlinear attachment}
\label{sec:seminonlinear}

Let us first consider the case $\alpha=1$ and $\beta\in \mathbb{R}$,
i.e. when the probability to attach to node $j$ at layer $1$ (resp.,
at layer $2$) is proportional to $k_jq_j^{\beta}$ (resp., to
$q_jk_j^{\beta}$).  In particular, when $\alpha=1$ and $\beta=0$, we
recover the uncorrelated linear preferential attachment kernel, which
has been extensively studied in Ref.~\cite{Nicosia2013}. In this case,
the degree distribution in each layer is a power law $P(k)\sim
k^{-\gamma}$ with $\gamma=3$, and the multiplex exhibits positive
inter-layer degree correlations, the degree of a node being
essentially determined by its age.

\begin{figure}[!t]
  \begin{center}
    \includegraphics[width=3in]{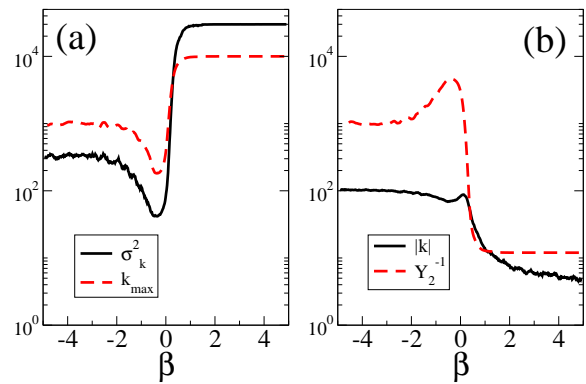}
  \end{center}
  \caption{(color online) Properties of the layer degree distribution
    for $\alpha=1.0$ as a function of $\beta$. (a) Variance
    $\sigma^2_k$ of the degree sequence (solid black) and maximum
    degree $k_{\rm max}$ (dashed red). (b) Number of different degree
    classes $|k|$ (solid black) and participation ratio $Y_2^{-1}$
    (dashed red). There is a clear dependence of the network structure
    on the attachment exponent $\beta$. The plots correspond to a
    multiplex network with $N=10.000$, $m=3$, $m_0=3$.}
  \label{fig:fig1_new}
\end{figure}

\begin{figure*}[!t]
  \begin{center}
    \includegraphics[width=6.2in]{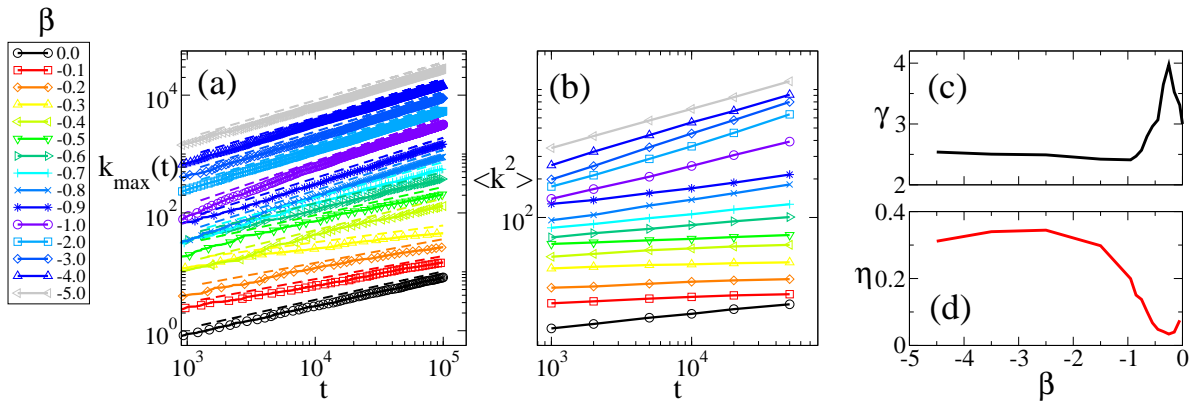}
  \end{center}
  \caption{(color online) The scalings with $t$ of the degree $k_{\rm
      max}(t)$ of the largest hub (a) and of the fluctuations of the
    degree sequence $\avg{k^2}$ (b) for $\alpha=1.0$ and for different
    values of $\beta\le 0$ suggest that the degree distribution of
    each layer is a power-law $P(k) \sim k^{-\gamma}$ (the plots are
    vertically displaced to enhance readability). (c) The exponent
    $\gamma$ of the degree distribution is equal to $3.0$ when
    $\beta=0$, as in the case of classical linear preferential
    attachment, has a maximum around $\beta\simeq -0.3$ and converges
    to $\gamma\simeq 2.5$ when $\beta\rightarrow -\infty$. (d)
    Similarly, the exponent $\eta$ has a minimum for $\beta~\simeq
    0.3$ and converges towards $\eta\simeq -0.3$ for negative values
    of $\beta$.}
  \label{fig:fig2_new}
\end{figure*}

When $\beta\neq 0$ the growth process can produce multiplex networks
with homogeneous, heterogeneous or condensed degree distribution on
each layer, characterized by either assortative or disassortative
inter-layer degree correlation patterns, depending on the sign of
$\beta$. In Fig.~\ref{fig:fig1_new} we report the results obtained by
simulating the growth of a multiplex for $\alpha=1.0$ and $\beta$ in
the range $[-5,5]$. In order to characterize the degree distributions
of the two layers we plot, as a function of $\beta$, the variance
$\sigma^2_{k}$ of the degree distribution, the maximum degree $k_{\rm
  max}$, the number $|k|$ of different degree classes present in each
layer, and the participation ratio $Y_2^{-1}$.  Given a degree
sequence $\{k_1, k_2, \ldots, k_N\}$, $Y_{2}^{-1}$ is defined as
\cite{Derrida:1987}
\begin{equation}
  Y_{2}^{-1} = \left[\sum_{i}\left(\frac{k_i}{\sum_j
      k_j}\right)^{2}\right]^{-1}.
\end{equation}

It is easy to show that $Y_{2}^{-1}\sim O(N)$ when $k_i=\avg{k}\>$ for
all $i$, i.e. for homogeneous degree distributions, while $Y_2^{-1}
= c \ll N$ if most values of $k_i$ are equal, except for a few nodes
for which we have $k_{i*}\simeq N$, i.e. in the presence of a
condensate state where a few nodes connect to nearly all the other
nodes of a layer. 
When $\alpha=1$ and $\beta$ is positive, we observe a transition to a
condensed state, characterized by small $|k|$, large $\sigma^2_k$,
$k_{\rm max}\sim O(N)$, and $Y_2^{-1} \sim O(1)$, signalling the
existence of a few dominant nodes.
Conversely, for negative values of $\beta$ we obtain heterogeneous
degree distributions (large values of $|k|$, relatively large values
of $k_{\rm max}$, $\sigma^2_k$ and $Y_2^{-1}$).  In particular, these
distribution are power-laws. In fact, as shown
in. Fig.~\ref{fig:fig2_new}(a)-(b), for $\beta\le 0$ the degree
$k_{\rm max}(t)$ of the largest hub of the graph at time $t$ scales as
$ t^{\varepsilon}, \quad \varepsilon>0$ and the fluctuations of the
degree distributions $\avg{k^2}$ scale as $t^{\eta}, \quad \eta
>0$. 

In Fig.~\ref{fig:fig2_new}(c) we report the exponent $\gamma$ of the
power law distribution of the two layers, whose value clearly depends
on $\beta$. In particular, for $\beta=0$ we recover $\gamma=3.0$, as
in the standard single-layer linear preferential attachment. When
$\beta\rightarrow -\infty$ then $\gamma$ converges to $\gamma\simeq
2.5$. When $\beta$ is negative and close to zero, we observe a strange
phenomenon, which is also responsible for the peaks in $|k|$,
$Y_2^{-1}$, $\sigma^2_k$ and $k_{\rm max}$ shown in
Fig.~\ref{fig:fig1_new}. In this region, as we increase $\beta$ the
distribution becomes first more homogeneous (with a peak of
$\gamma\simeq 4$ for $\beta\simeq -0.3$) and then again more
heterogeneous, up until $\gamma=3.0$ for $\beta=0$.

This apparently strange behavior can be explained by considering that
for $\beta<0$ the two layers are competing, i.e. a node having high
degree on one layer will tend to have small degree on the other
layer. In this case, a small negative value of $\beta$ actually
reduces the heterogeneity of the attachment probability distribution
that we have for $\beta=0$, allowing small-degree nodes (for which the
effect of layer competition is mitigated by the fact that $\beta$ is
negative and close to zero) to acquire more edges. Conversely, when
the value of $\beta$ becomes smaller then local fluctuations start to
play a fundamental role, and the distribution becomes more
heterogeneous again. 

\begin{figure*}[!t]
  \begin{center}
    \includegraphics[width=6.2in]{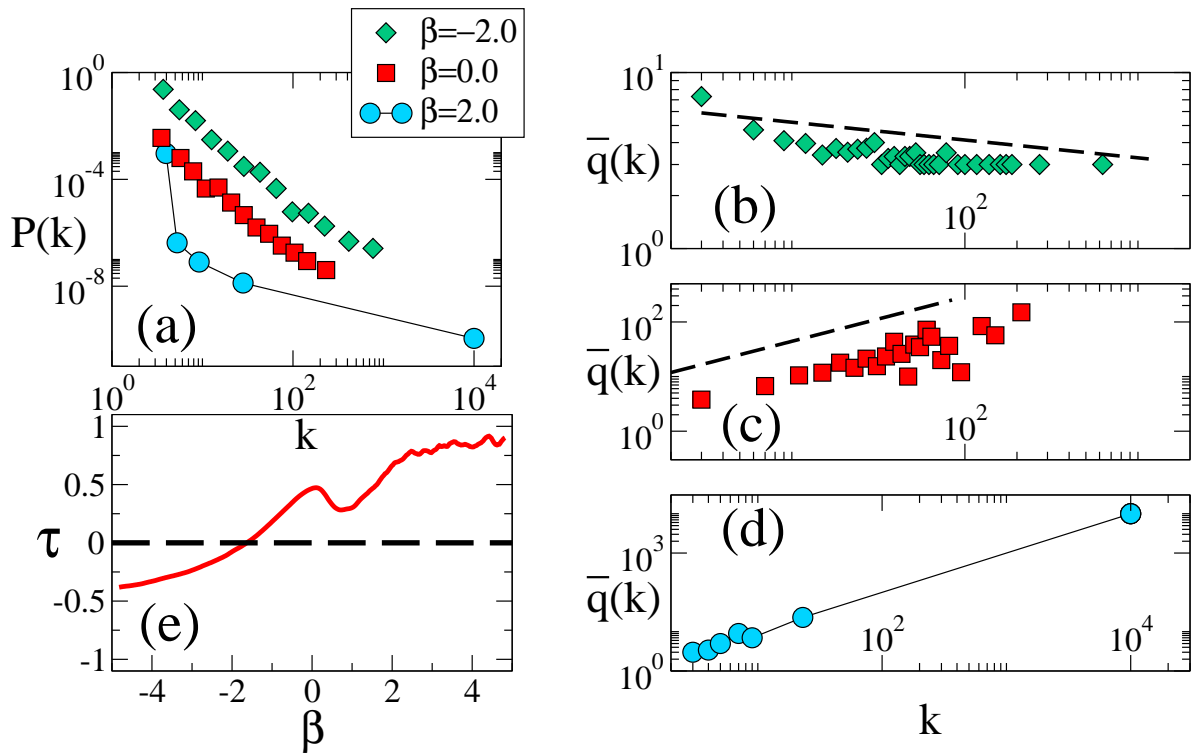}
  \end{center}
  \caption{(color online) (a) Typical degree distributions for
    $\beta=-2.0$ (green diamonds), $\beta=0.0$ (red squares) and
    $\beta=2.0$ (cyan circles) when $\alpha=1.0$. According to the
    value of $\beta$, the degree distribution of a layer can be either
    a power-law ($\beta \le 0$) or condensed, i.e. characterized by
    the presence of a super-hub which attracts an extensive fraction
    of edges.  The corresponding pattern of inter-layer degree
    correlations $\overline{q}(k)$ in the three cases [panels (b)-(d)]
    and the plot of the inter-layer degree correlation coefficient
    $\tau$ [Panel (e)] confirm that both positive and negative
    correlations can be obtained by tuning $\beta$.}
  \label{fig:fig3_new}
\end{figure*}

Three typical examples of degree distributions obtained for different
values of $\beta$ are reported in Fig.\ref{fig:fig3_new}(a), in
particular for $\beta= -2.0, 0.0, 2.0$. When $\beta=2.0$ we observe a
homogeneous distribution for small values of $k$, and one node
acquires a finite fraction of the edges, i.e. the network is
condensed. For $\beta=0.0$ the degree distribution is a power-law with
exponent $\gamma\simeq 3.0$~\cite{Nicosia2013}. Finally, for
$\beta=-2.0$ the degree distribution can be fitted by a power-law with
exponent $\gamma\simeq 2.5$.

Interestingly, the value of $\beta$ also determines the sign and value
of inter-layer degree correlations, as confirmed by the plot of the
average degree $\overline{q}(k)$, at layer $2$, of nodes having degree
$k$ at layer $1$, shown in
Fig.~\ref{fig:fig3_new}(b)-(d)~\cite{Nicosia2013}. It is clear that,
by tuning $\beta$, one can obtain either positive ($\beta=2.0$, and
$\beta=0.0$) or negative ($\beta=-2$) inter-layer degree correlations.
Finally, in Fig.~\ref{fig:fig3_new}(e) we plot as a function of
$\beta$ the value of the Kendall's rank correlation coefficient $\tau$
computed on the degree sequences of the two layers (see Appendix). For
$\beta<0$ we have disassortative inter-layer degree correlations
($\tau<0$), meaning that a hub on one layer is a poorly-connected node
on the other layer, while for $\beta>0$ the degrees of the two
replicas of the same node are positively correlated ($\tau>0$).

\section{Mean-field approximation}
\label{sec:meanfield}

The mean-field approach has been proven to be extremely good in making
qualitative predictions on the degree distribution of growing network
models. Here we report convincing evidence that this approach is not
able to capture essential properties of the proposed non-linear growth
model. In fact in this model stochastic effects are fundamental to
describe the evolution of the system.  As we will see in a moment, the
conclusion of the mean-field theory is that the expected degrees of
the same node in the two layers are equal, irrespective of the values
of the two exponents $\alpha$ and $\beta$. However, such a conclusion
is in clear disagreement with the results obtained by numerical
simulation and reported, for instance, in Fig.~\ref{fig:fig3_new},
which indeed confirm that for some combinations of the exponents
$\alpha$ and $\beta$ the degrees of a node at the two layers can be
negatively correlated.

In the mean-field approximation the degree of a node $i$ at time $t$
acquires a deterministic value equal to its average degree in the
stochastic model.  If we indicate by $\kappa_i(t)$ and by $\phi_i(t)$
the average degree of node $i$ on layer 1 and on layer 2 respectively,
the mean-field approximation assumes that the degree $k_i(t)$ of a
node $i$ in layer 1 is equal to $\kappa_i(t)$,
i.e. $k_i(t)=\kappa_i(t)$ and similarly that the degree in layer 2
$q_i(t)$ of node $i$ at time $t$ is given by $q_i(t)=\phi_i(t)$.
Since, in this approximation, the average number of links that at time
$t$ a node $i$ acquires in layer 1 is given by
\begin{equation}
m\frac{f(\kappa_i,\phi_i)}{\sum_j f(\kappa_j,\phi_j)},
\end{equation}
while the average number of links  that a node $i$  acquires at time
$t$ in layer 2  is given by 
\begin{equation}
m\frac{f(\phi_i,\kappa_i)}{\sum_j f(\phi_j,\kappa_j)},
\end{equation}
when $f(\kappa,\phi)=\kappa^{\alpha}\phi^{\beta}$ and
$f(\phi,\kappa)=\phi^{\alpha}\kappa^{\beta}$ as in Eq.(3), the
mean-field equations for $\kappa_i(t)$ and $\phi_i(t)$ at large times
$t\gg 1$ read
\begin{equation}
\frac{d\kappa_i}{dt}=\frac{\kappa_i^{\alpha}\phi_i^{\beta}}{Ct},~~~~~~~~~~~~~\frac{d
  \phi_i}{dt}=\frac{\phi_i^{\alpha} \kappa_i^{\beta}}{Ct},
\label{mf}
\end{equation}
with the constant $C$ to be self-consistently determined as
\begin{equation}
C=\lim_{t \to \infty}\frac{\sum_{i=1}^t {\kappa_i^{\alpha}\phi_i^{\beta}}}{mt}.
\end{equation}
Assuming that $C$ is a constant, the Eqs. $(\ref{mf})$ can be
rewritten as
\begin{equation}
\kappa_i^{\beta-\alpha}\frac{d \kappa_i}{d \ln
  t}=\frac{1}{C}{(\kappa_i\phi_i)^{\beta}}=\phi_i^{\beta-\alpha}\frac{d
  \phi_i}{d \ln t}.
\label{mf0}
\end{equation}
Therefore we find for $\beta-\alpha\neq -1$
\begin{equation}
  \frac{d  \left[\kappa_i^{\beta-\alpha+1}-\phi_i^{\beta-\alpha+1}\right]}{d \ln t}=0, 
\label{mf1}
\end{equation}
while we have for $\beta-\alpha=-1$
\begin{equation}
\frac{d \ln \kappa_i-\ln \phi_i}{d \ln t}=0.
\label{mf2}
\end{equation}
Therefore, if we consider the initial conditions
$\kappa_i(t_i)=\phi_i(t_i)=m$ the mean-field approach implies always
$\kappa_i(t)=\phi_i(t)$.  Inserting this solutions in the
Eqs. $(\ref{mf})$ we get
\begin{equation}
\frac{d\kappa_i}{dt}=\frac{\kappa_i^{\alpha+\beta}}{Ct}
\end{equation}
yielding the solution
\begin{equation}
\kappa_i(t)=m\left(\frac{t}{t_i}\right)^{1/C}
\label{mf4a}
\end{equation}
for $\alpha+\beta=1$, and  the solution

\begin{equation}
\kappa_i(t)=\left[m^{1-(\alpha+\beta)}+\frac{1-\alpha-\beta}{C}\ln\left(\frac{t}{t_i}\right)\right]^{1/(1-(\alpha+\beta))}
\label{mf4b}
\end{equation}
for $\alpha+\beta< 1$.

For $\alpha+\beta>1$ we observe a singularity in the solution for
$\kappa_i(t)$ indicating the fact that the self-consistent equation
for $C$ cannot be satisfied. By studying the master equation we will
show that for $\alpha+\beta>1$ we observe a condensation phase
transition.  Starting form the solution given by Eq.~(\ref{mf4a}) and
Eq.(\ref{mf4b}) the predicted degree distribution is scale free with
power-law exponent $\gamma=1+1/C$ for $\alpha+\beta=1$ and a Weibull
distribution for $\alpha+\beta<1$.

Overall we can say that the mean-field approach provides a solution
that reflects the symmetry of the model in the two
layers. Nevertheless this approach mostly fails in characterizing the
correlations between the degrees of the same node in different
layers. As we said before, the behavior of the model and its
predictions $\kappa_i(t)=\phi_i(t)$ are not supported by the
simulations because the dynamics of the model is strongly affected by
stochasticity and noise. In particular we argue that the strong
deviations from the mean-field behavior that we observe in the
simulations are due to the fundamental role played by stochastic
effects on the degrees of the nodes recently arrived in the
network. In fact these nodes will have a small degree in both layers
and the fluctuations on these quantities will strongly affect the
linking probability distribution.

\section{Master equation}
\label{sec:master_eq}

More theoretical insights about the model defined by Eq.~(\ref{pi})
and Eq.~(\ref{f}) come from the solution of the master equation of the
system, which accounts for the expected number of nodes $N_{k,q}$ with
$k$ links in layer 1 and $q$ links in layer 2. Let us consider for
simplicity the case $m=1$. The master equation needs to take into
account that at any time $t$ one of the following events can occur:

\begin{enumerate}[i) ]
\item The number of nodes $N_{k,q}$ with degree $k$ in
  layer 1 and degree $q$ in layer 2 increases by one if the new node
  links in layer 1 but not in layer 2 to a node of degree $k-1$ in
  layer 1 and degree $q$ in layer 2.
\item The number of nodes $N_{k,q}$ with degree $k$
  in layer 1 and degree $q$ in layer 2 increases by one if the new
  node links in layer 2 but not in layer 1 to a node of degree $k$ in
  layer 1 and degree $q-1$ in layer 2.
\item The number of nodes $N_{k,q}$ with degree $k$
  in layer 1 and degree $q$ in layer 2 increases by one if the new
  node links in layer 1 and also in layer 2 to a node of degree $k-1$
  in layer 1 and degree $q-1$ in layer 2.
\item The number of nodes $N_{k,q}$ with degree $k$
  in layer 1 and degree $q$ in layer 2 decreases by one if the new
  node links in layer 1 to a node of degree $k$ in layer 1 and degree
  $q$ in layer 2.
\item The number of nodes $N_{k,q}$ with degree $k$ in
  layer 1 and degree $q$ in layer 2 decreases by one if the new node
  links in layer 2 to a node of degree $k$ in layer 1 and degree $q$
  in layer 2.
\end{enumerate}
Moreover, for $k=m$ and $q=m$ the average number of nodes $N_{k,q}$
with degree $k$ in layer 1 and degree $q$ in layer 2 increases by one
at each time step, since the newly arrived node has degrees $k=m$ and
$q=m$. Taking into account all these possibilities, we can write the
master equation as

\begin{widetext}
\begin{eqnarray}
N_{k,q}(t+1) &=& N_{k,q}(t) + \delta_{k,m}\delta_{q,m} +\frac{f(k-1,q)}{{\cal
    M}(t)}\left(1-\frac{f(q,k-1)}{{\cal
    M}(t)}\right)N_{k-1,q}(t)(1-\delta_{k,m})\nonumber \\
    &&+\frac{f(q-1,k)}{{\cal M}(t)}\left(1-\frac{f(k,q-1)}{{\cal
    M}(t)}\right)N_{k,q-1}(t)(1-\delta_{q,m})  -\left[\frac{f(q,k)+f(k,q)}{{\cal M}(t)}-\frac{f(k,q)f(q,k)}  {[{\cal M}(t)]^2  }\right]{N_{k,q}(t)}  \nonumber \\
 &&+ \frac{f(q-1,k-1)f(k-1,q-1)}{[{\cal M}(t)]^2}N_{k-1,q-1}(t)(1-\delta_{q,m})(1-\delta_{k,m})
\label{rate2}
\end{eqnarray} 
\end{widetext}
where $k,q \ge m$, $f(k,q)=k^{\alpha}q^{\beta}$,
$\delta_{\bullet,\bullet}$ is the Kronecker delta and ${\cal M}(t)$ is
given by:
\begin{equation} {\cal M}(t)=\sum_{k,q}
f(k,q)N_{k,q}(t)=\sum_{k,q} f(q,k)N_{k,q}(t)
\end{equation}

Eq.~(\ref{rate2}) can be solved by using techniques similar to those
adopted for single-layer networks or for multiplex networks with
linear or semi-linear attachment
kernels~\cite{krapivsky,Dorogovtsev,Nicosia2013}. In particular, by
solving the master equation we obtain an analytical explanation for
the appearance of a condensed phase. In fact, the master equation
depends on the quantity ${\cal M}(t)$ which satisfies, in the
thermodynamic limit $t\to \infty$, the relation
\begin{equation} {\cal M}(t)=\sum_{k,q}
f(k,q)N_{k,q}(t)=\sum_{k,q} f(q,k)N_{k,q}(t).
\end{equation} 
Assuming that the normalization sum scales like ${\cal M}(t)\propto
t$, i.e.  $\lim_{t\rightarrow\infty}{{\cal M}(t)/t}=C$ with $C$
constant, we can derive a recursive expression for $P_{k,q}=\lim_{t\to
  \infty}N_{k,q}(t)/t$.  However, the hypothesis ${\cal M}(t)\propto
t$ depends on the value of the exponents $(\alpha,\beta)$ and in
general is not satisfied.  A deviation from this scaling indicates
that in each layer we have a node that is grabbing an extensive number
of links $k_{\rm max}\simeq t,q_{\rm max}\simeq t$, i.e.  we are in a
condensed network phase.

\subsection{Conditions for condensation}

In order to show in which region of the phase space condensation
occurs we first find a sufficient condition for condensation and then
we will show that this condition is also a necessary one. We make use
of the master equation to estimate ${\cal M}(t)$, respectively for
$\beta\leq 0$ and $\beta>0$, by considering (without loss of
generality) the case $m=1$. We observe that for $\beta\leq 0$ at each
time $t$ there are no vertices that in layer 1 have degree greater
than $k=t$, therefore the master equation given by Eq.~(\ref{rate2})
becomes
\begin{equation}
  N_{k,1}(k)=\frac{(k-1)^{\alpha}}{{\cal
      M}(k-1)}\left(1-\frac{(k-1)^{\beta}}{{\cal
      M}(k)}\right)N_{k-1,1}(k-1) \end{equation}
But for large times $\frac{(k-1)^{\beta}}{{\cal M}(k)}\ll 1$. Moreover
the fractions $N_{k-1,1}(k-1)/N_{k,1}(k)\geq1$ since only the first
node of the network can have degree $k$ equal to the time $t=k$.
Consequently 
\begin{equation} {\cal
    M}(t)\geq{t^{\alpha}}.  \end{equation}
Instead, if $\beta>0$ then at time $t$ there are no nodes that have at
the same time degree in layer 1 greater than $k=t$ and degree in layer
2 greater then $k=t$. In this case the master equation given by
Eq.~(\ref{rate2}) becomes
\begin{equation} N_{k,k}(k)=\frac{(k-1)^{2(\alpha+\beta)}}{[{\cal
        M}(k-1)]^2}N_{k-1,k-1}(k-1).  
\end{equation}
\begin{figure*}[!t]
  \begin{center}
    \includegraphics[width=7in]{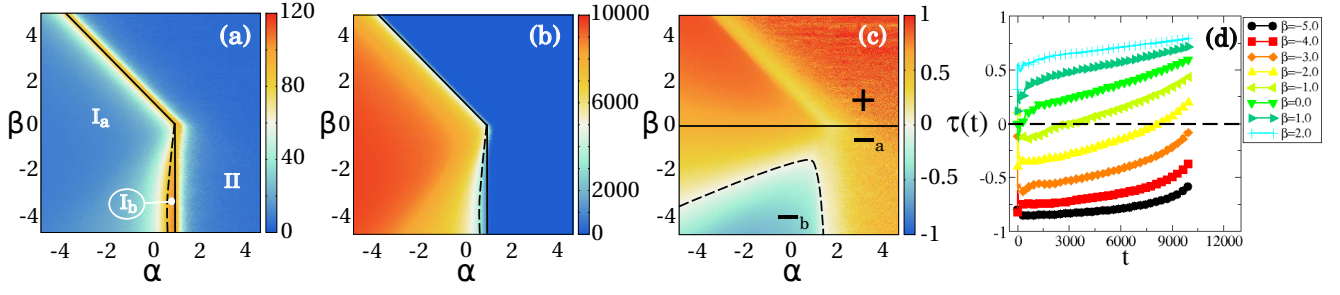}
  \end{center}
  \caption{(color online) As a function of the two parameters $\alpha$
    and $\beta$ we report, by means of a color code: (a) the number of
    distinct degree classes $|k|$, (b) the participation ratio
    $Y_2^{-1}$, and (c) the Kendall's $\tau$ correlation
    coefficient. The solid black lines in panel (a) and (b) separate
    the non-condensed (region I) from the condensed phase (region II,
    small $|k|$, small $Y_2^{-1}$). In region I we can have either
    homogeneous (region ${\rm I_a}$) or heterogeneous degree
    distributions (region ${\rm I_b}$). The solid black line in panel
    (c) separates the two regions with positive (region $\bm{+}$) and
    negative inter-layer degree correlations (region $\bm{-}$),
    respectively corresponding to $\beta>0$ and $\beta<0$. The value
    of $\tau$ for the whole multiplex is negative only in region
    $\bm{-_b}$. In panel (d) we show the plot of $\tau(t)$, which is
    the Kendall's $\tau$ restricted to the nodes arrived up to time
    $t$. The dashed black line corresponds to $\tau=0$ and is reported
    for visual reference. }
  \label{fig:fig4_new}
\end{figure*}
Notice that the fractions $N_{k-1,k-1}(k-1)/N_{k,k}(k)\geq1$ since
only the first node of the network can have degrees $(k,q)$ equal to
$(t,t)$, where $t$ is the time. Therefore we get that 
\begin{equation}
{\cal M}(t)\geq\left\{\begin{array}{ccc}t^{\alpha}  & \mbox{if} & \beta\leq 0 \nonumber \\
t^{\alpha+\beta}&\mbox{if} & \beta> 0\end{array}\right.
\end{equation}
This means that for $\alpha>1$ and $\beta<0$ or for $\beta>0$ and
$\alpha> 1-\beta$ 
\begin{equation} 
  {\cal M}(t)\geq t^{\xi}, \quad\forall  \xi>1.
  \label{sc}
\end{equation} 
This is a sufficient condition to have condensation, since in this
case the expected number of nodes that at time $t$ have degrees
$k=1,q=1$ scales with $t$, i.e. $N_{1,1}(t)\simeq t$.  In fact,
starting from the master equation, $N_{1,1}(t)$ satisfies the
following relation
\begin{equation} \frac{dN_{1,1}(t)}{dt}=-2\frac{1}{{\cal
      M}(t)}N_{1,1}(t)+1,
\label{11}
\end{equation} 
where in writing this equation we have neglected higher order terms in
$[{\cal M}(t)]^{-1}$.  If Eq.~(\ref{sc}) is satisfied, then the first
term in the right-hand side of Eq.~(\ref{11}) is negligible and we
have $N_{1,1}(t)\simeq t$ for large $t$. This implies that the number
of nodes with degrees different from $(k=1,q=1)$ is negligible, so
that in this region we have a condensation phenomenon with few nodes
grabbing an extensive number of connections.

Let us now show that the condition $\beta<0$, $\alpha>1$ and
$\beta>0$, $\alpha+\beta>1$ is also necessary for condensation.  Let
us assume that we have a condensation of the links. In this scenario,
we will have for $\beta<0$ one node with degree $k=t$ on layer 1, say
node $i$, and another node with degree $q=t$ on layer 2, say node $j$;
conversely, for $\beta>0$ we will expect to have exactly one node, say
node $i$, having degrees $(k,q)=(t,t)$.  Since we have condensation
then we can write an upper bound to ${\cal
  M}(t)=\sum_{k,q}k^{\alpha}q^{\beta}N_{k,q}$, by taking into account
only the contribution of the condensed nodes:
\begin{equation} {\cal M}(t)\leq
  \left\{\begin{array}{ccc}t^{\alpha}&\mbox{for}& \beta\leq 0
  \\ t^{\alpha+\beta}&\mbox{for}& \beta> 0\end{array}\right.
\label{lb}
\end{equation}
Putting Eq.~(\ref{lb}) together with the lower bound given by
Eq.~(\ref{sc}) we find that ${\cal M}(t)$ satisfies the scaling
\begin{equation} {\cal M}(t)\simeq
  \left\{\begin{array}{ccc}t^{\alpha}&\mbox{for}& \beta\leq
  0\nonumber\\ t^{\alpha+\beta}&\mbox{for}&\beta<0\end{array}\right.
\end{equation} 
But we know that ${\cal M}(t)\propto t^{\xi}$ with $\xi \geq 1$,
therefore we confirm that if the condensation transition occurs then
either $\alpha>1$ and $\beta\leq 0$ or $\alpha+\beta>1$ and
$\beta>0$. Therefore the condensation transition occurs only in the
region $\beta<0$ $\alpha>1$ or in the region $\beta>0$,
$\alpha>1-\beta$.  In particular, for $\beta>0$ the same node will be
the condensate node in both layers, while for $\beta<0$ the condensate
node in one layer will not be the condensate node in the other
layer. When $\beta=0$ the condensate nodes in the two layers might be
either the same node or different nodes in different realizations.

\subsection{Solution of the master equation in the non
  condensed phase}

We consider now the master equation in the non condensed phase where
${\cal M}(t)\simeq Ct$ with $C>0$ independent on $t$, for $t\gg 1$.
As we have seen above, this implies that the parameters $\alpha,\beta$
satisfy the conditions: $\alpha\leq 1$ and $\beta<0$ or $\beta>0$ and
$\alpha \leq 1-\beta$.  In this region of the phase space, we have
always $f(k,q)/{\cal M}(t) \ll 1$ and therefore we can neglect the
terms proportional to $[{\cal M}(t)]^{-2}$ in the rate equation,
finding the master equation for evolving multiplex in the non
condensed phase, i.e.
\begin{eqnarray}
  \frac{dN_{k,q}(t)}{dt} & = 
  \frac{  \displaystyle A_{k-1,q}}{\displaystyle t}N_{k-1,q}(t)+\frac{
    \displaystyle B_{k,q-1}}{  \displaystyle t}N_{k,q-1}(t) +\nonumber
  \\
  &-\left[\frac{  \displaystyle A_{k,q} + B_{k,q}}{  \displaystyle t}\right]N_{k,q}(t)+\delta_{k,m}\delta_{q,m}
\end{eqnarray} where we have put 
\begin{equation} 
A_{k,q}=\frac{k^{\alpha}q^{\beta}}{C
},~~~~~~~~~~~~~~~
B_{k,q}=\frac{q^{\alpha}k^{\beta}}{C
} ,\end{equation} 
and $C$ is a constant that can be determined self-consistently as
\begin{equation} C=\lim_{t\to \infty}\frac{1}{t}{\sum_{k,q}
    k^{\alpha}q^{\beta}N_{k,q}}(t).
\end{equation} 

Assuming $N_{k,q}\simeq tP_{k,q}$ valid in the large time limit, we
can solve for $P_{k,q}$ and we get 
\begin{equation}
\begin{array}{cll}
P_{m,q}&=&\left(\prod_{j=m}^{q}\frac{B_{k,j-1}}{1+A_{k,j}+B_{k,j}}\right)P_{m,m}\nonumber
\\ P_{k,q}&=&\sum_{r=1}^{q}\left(\prod_{j=r+1}^{q}
\frac{B_{k,j-1}}{1+A_{k,j}+B_{k,j}}\right)\frac{A_{k-1,r}}{1+A_{k,r}+B_{k,r}}P_{k-1,r}
\end{array}
\end{equation}
These recursive equations can be used to solve numerically for the
joint degree distribution of the degrees in the two layers, but
unfortunately for $\beta\neq 0$ there is no closed form analytical
solution to these equations.

\begin{figure*}[!t]
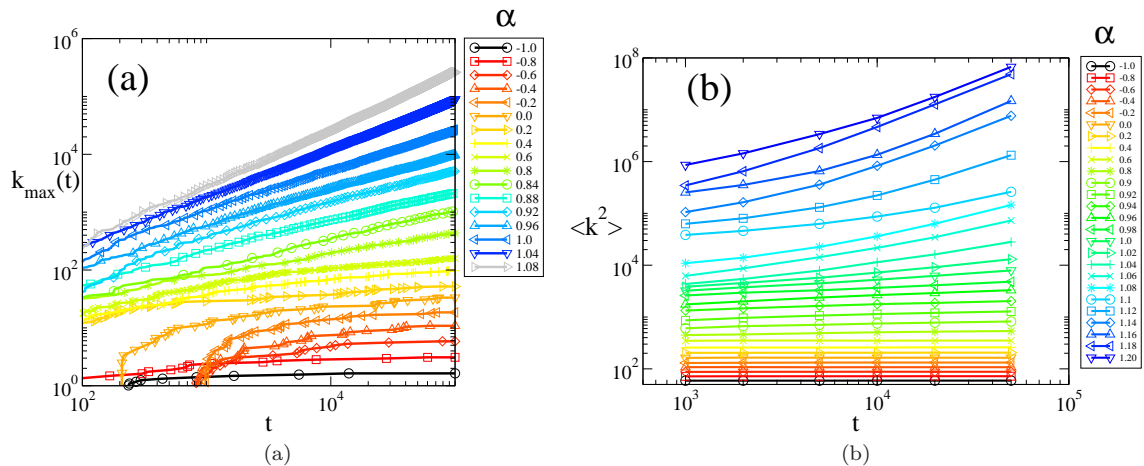

  \begin{center}
    \subfigure[]{
      \includegraphics[width=2.8in]{fig5a_new.eps} }
    \subfigure[]{
      \includegraphics[width=3in]{fig5b_new.eps} }
  \end{center}
  \caption{(a) The scaling of the degree $k_{\rm max}(t)$ of the
    largest hub of a layer depends on the value of $\alpha$ (here we
    fixed $\beta=-1.0$). In particular, $k_{\rm max}(t)\sim
    t^{\varepsilon}$ only for $\alpha$ larger than $0.6\sim 0.8$,
    suggesting that the degree distribution becomes heterogeneous when
    $\alpha$ is closer to the critical value for condensation
    ($\alpha=1.0$). (b) The value of $\avg{k^2}$ as a function of $t$,
    for $\beta=-1.0$ and different values of $\alpha$. If we start
    from $\alpha=-1.0$ and keep increasing it, we initially notice no
    scaling at all with $t$, up until $\alpha\simeq 0.8$, when
    $\avg{k^2}\sim t^{\eta}$. This means that for $0.8<\alpha<1.0$ the
    second moment of the degree distribution diverges with $t$.}
  \label{fig:cond_trans}
\end{figure*}

\section{Numerical Results}
\label{sec:numerical}

The predictions obtained by solving the master equation of the model
are in very good agreement with the phase diagram of the system
obtained through simulations, reported in
Fig.~\ref{fig:fig4_new}(a)-(b) ($N=10.000$, $m=3$, $m_0=3$). In these
figures we show, for each value of the two parameters $\alpha$ and
$\beta$, the corresponding values of $|k|$ (a) and $Y_2^{-1}$ (b),
which allow us to visualize the two separate regions of the phase
space. In region I the degree distribution is not condensed, while in
region II we observe condensation as indicated by both a small value
of $|k|$ and of $Y_2^{-1}$. The shape of the boundary between the two
regions agrees very well with the analytical prediction provided by
the solution of the master equation in the thermodynamic limit
(indicated by the solid lines in panel (a) and panel (b)). We notice
that region I can be further divided into two separate sub-regions,
according to the fact that the resulting degree distribution at each
layer is homogeneous (region ${\rm I_a}$) or heterogeneous (region
${\rm I_b}$).

\begin{figure*}[!t]
  \begin{center}
    \includegraphics[width=6.2in]{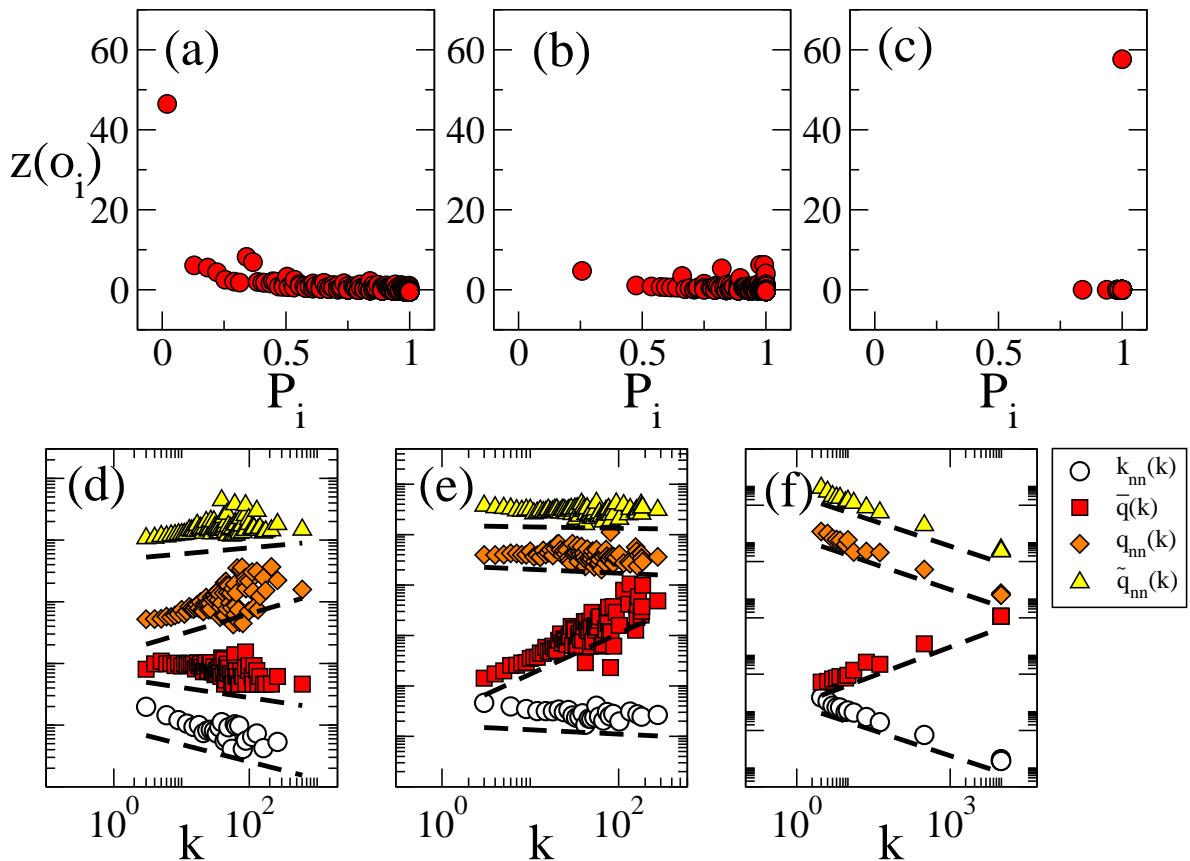}
  \end{center}
  \caption{(color online) In the top panels we report the multiplex
    cartography of networks obtained by setting, respectively, (a)
    $\beta=-1.0$, (b) $\beta=0.0$ and (c) $\beta=1.0$ when
    $\alpha=1.0$. For $\beta < 0$ there is a high heterogeneity of
    node roles, and hubs tend to be focused. Conversely, for $\beta>0$
    all nodes tend to be genuinely multiplex, i.e. to have similar
    degrees on both layers. In the bottom panels we plot the
    intra-layer, inter-layer and mixed correlations respectively for
    (d) $\beta=-1.0$, (e) $\beta=0.0$ and (f) $\beta=1.0$ when
    $\alpha=1.0$.}
  \label{fig:fig7_new}
\end{figure*}

It is interesting to analyze the transition to condensation as a
function of $\alpha$ at fixed $\beta<0$, i.e. region $I_{b}$. In
particular, we are interested in checking whether the degree
distribution becomes a power--law before we reach the condensation
transition (we already know that at the boundary of the condensation
transition the degree distribution is a power-law, with an exponent
which depends on $\beta$, as discussed in
Sec.~\ref{sec:seminonlinear}). Therefore, we analyzed the scaling of
the degree of the largest hub $k_{\rm max}(t)$ and of the fluctuations
of the degree distribution $\avg{k^2}$ as a for increasing values of
$\alpha$. The results corresponding to $\beta=-1.0$ are reported in
Fig.\ref{fig:cond_trans}(a) and Fig.~\ref{fig:cond_trans}(b). Notice
that for $\alpha<0 $ we observe homogeneous degree distributions,
i.e. no scaling of fluctuations with $N$ and a logarithmic scaling of
$k_{\rm max}(t)$, while for $\alpha=1.0$ we have $k_{\rm max}(t) \sim
t^{1/2}$, which corresponds to $\gamma=3.0$. However, we observe that
$\avg{k^2}$ scales as $t^{\eta}$ already for $\alpha<1.0$, and in
particular in the region $0.8<\alpha<1.0$. Also, in this region
$k_{\rm max}(t)$ scales as $t^{\varepsilon}$, indicating that in
region $I_b$ the degree distribution of each layer is a power-law.

Concerning the sign of inter-layer correlations, in
Fig.~\ref{fig:fig4_new}(c) we report the Kendall's correlation
coefficient $\tau$ of the degree sequences at the two layers, where
the two regions where inter-layer degree correlations are respectively
positive (region $\bm{+}$) and negative (region $\bm{-}$) are
separated by a solid black line. It is interesting to note that a
multiplex can exhibit either positive or negative inter-layer
correlations independently of the fact that its layers have
homogeneous or heterogeneous distributions. While from the linking
probabilities given by Eq.~(\ref{pi})-(\ref{f}) we expect $\tau>0$
when $\beta>0$, when $\beta<0$ the degrees of a node in the two layers
tend to be negatively correlated. However, the interpretation of the
phase diagram of $\tau$ for negative $\beta$ is less trivial, and the
shape of the boundary between the regions $\bm{-_a}$ and $\bm{-_b}$
needs some explanation. In fact, when $\beta<0$ Eq.~(\ref{f}) implies
that if a node has high degree in one layer, it will have low
probability to acquire new links in the other layer, so that the
degrees of the old nodes of the network will be negatively correlated.
This is clear by looking at Fig.~\ref{fig:fig4_new}(d), which confirms
that for $\beta<0$ the inter-layer degree correlations of older nodes
are always negative.

However, for some values of $\beta$ the value of $\tau$ computed on
the whole network could be positive, due to the presence of a large
majority of younger nodes having small degrees on both layers
(i.e. fickle nodes), whose values are mostly determined by stochastic
fluctuations.  In general, for large negative values of $\beta$ the
fraction of fickle nodes is reduced, until it becomes zero for
$\beta<\beta_c(\alpha)$ (the dashed line in Fig.~\ref{fig:fig4_new}(c)
corresponds to the values of $\beta_c(\alpha)$), and in this case all
the nodes have negative correlated degrees, resulting in a negative
value of $\tau$. We notice that the existence of two sub-regions in
the phase diagram of $\tau$ for $\beta<0$ is not a finite-size effect,
as confirmed by the results shown in Appendix
Fig.~\ref{fig:tau_sizedep}.

\section{Multiplex cartography}
\label{sec:cartography}

The authors of Ref.~\cite{battiston} have recently introduced the
concept of multiplex cartography, which is in the same spirit of the
network cartography proposed by Guimer\'a and Amaral in
Ref.~\cite{Guimera05,GuimeraNature05}. Multiplex cartography is based
on two measures, namely the Z-score of the overlapping degree of a
node:
\begin{equation}
  z(o_i)=\frac{o_i- \langle o \rangle}{\sigma_o}
\end{equation}
where $o_i = \sum_\alpha k\lay{\alpha}_i$ while $\avg{o}$ and
$\sigma_o$ are the average and standard deviation of $o_i$ over all
the nodes, and the multiplex participation coefficient:
\begin{equation}
  P_i=\frac{M}{M-1}\left[1-
    \sum_{\alpha=1}^M\left(\frac{k_i^{[\alpha]}}{o_i}\right)^2\right].
\end{equation}
The multiplex participation coefficient of a node characterizes its
involvement in the layers of the multiplex. In fact, $P_i$ tends to
$1$ if node $i$ has exactly the same degree on all the $M$ layers,
while $P_i=0$ if node $i$ is isolated on all the $M$ layers but one.
With respect to the Z-score of their overlapping degree, we
distinguish \textit{hubs}, for which $z(o_i)\ge 2$, from regular
nodes, for which $z(o_i)<2$. With respect to the multiplex
participation coefficient, we call \textit{focused} those nodes for
which $0\le P_i\le 0.3$, \textit{mixed} the nodes having $0.3<P_i\le
0.6$ and \textit{truly multiplex} (or even simply \textit{multiplex})
the nodes for which $P_i > 0.6$. The scatter-plot of $z(o_i)$ and
$P_i$ provides information about the patterns of participation across
nodes of different degree classes, and gives insight about the
different roles played by nodes.

In Fig.~\ref{fig:fig7_new}(a)-(c) we report the multiplex cartography
diagrams for different values of $\beta$ ($\alpha=1.0$).  It is
interesting to notice that layer competition (i.e., $\beta < 0$)
enhances the variability of the multiplex cartography but produces
multiplexes in which hubs are predominantly focused (top-left corner
of the plots) while poorly-connected nodes are predominantly
multiplex. Conversely, strong layer concordance (i.e., $\beta>0$)
tends to produce multiplexes in which nodes belong to just a few
different classes, i.e. either multiplex hubs or multiplex nodes.

\section{Mixed correlations}
\label{sec:mixed_corr}

Since the combination of $\alpha$ and $\beta$ allows to produce
multiplex graphs having either assortative or disassortative
intra-layer degree-degree correlations and positive, null or negative
inter-layer degree correlations, it is interesting to look at the
combination of intra-layer and inter-layer correlations. In
particular, we might ask whether a node being a hub on layer $1$ is
preferentially connected on layer $2$ with other hubs or instead with
leaves. So in general we can be interested in assessing whether:

\begin{enumerate}[i) ]
\item a hub tends to be connected with other hubs or to
  poorly-connected nodes (intra-layer correlations);
\item
   a hub on one layer tends to be either a hub or a poorly-connected
   node in the other layer (inter-layer correlations);
\item
  a hub in one layer has neighbors in the other layer who are
  connected either to other hubs or poorly-connected nodes(type-1
  mixed correlations).
\item
   the neighbors of a hub in one layer are either hubs or
   poorly-connected nodes in the other layer (type-2 mixed
   correlations).
\end{enumerate}

\begin{figure*}[!ht]
  \begin{center}
    \includegraphics[width=6in]{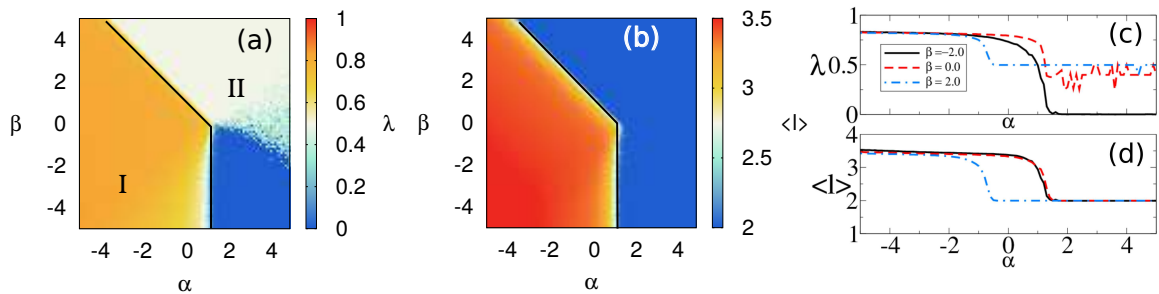}
  \end{center}
  \caption{(color online) Phase diagrams of (a) the average shortest
    path length $\avg{l}$ and (b) the multiplex interdependence
    $\lambda$ for $N=2000$, $m=3$, $m_0=3$, and three corresponding
    cross-cut sections for fixed values of $\beta$ [panel (c) and
      panel (d)]. As in Fig.~\ref{fig:fig4_new}, the black solid lines
    separate the non condensed (region $\rm I$) from the condensed
    phase (region $\rm II$). In the non-condensed phase both $\avg{l}$
    and $\lambda$ exhibit small variations, while the interdependence
    in the condensed phase is $\lambda=0.5$ for $\beta>0$ and
    $\lambda=0.0$ for $\beta<0$, in agreement with the fact that for
    $\beta>0$ the same node is condensed in both layers, while for
    $\beta<0$ the condensed nodes in the two layers are distinct.}
  \label{fig:fig8_new}
\end{figure*}

We measure type-1 mixed correlations using the quantity:
\begin{equation}
  k_{nn}(q) = \sum_{k} P(k=k_i|q=q_i) \frac{1}{k_i}
  \sum_{j}a\lay{1}_{ij} k_{j}
\end{equation}
which is the average degree of first neighbors on layer $1$ of a node
having degree $q$ at layer $2$. Similarly, we can define the
dual quantity:
\begin{equation}
  q_{nn}(k) = \sum_q P(q=q_i|k=k_i) \frac{1}{q_i}
  \sum_{j}a\lay{2}_{ij} q_{j}
\end{equation}
If the plot of $k_{nn}(q)$ is an increasing (decreasing) function of
$q$, then we say that the mixed correlations of layer $1$ with respect
to layer $2$ are positive (negative), or assortative (disassortative).

Type-2 mixed correlations can be quantified through the following
expression:
\begin{equation}
\widetilde{q}_{nn}(k)=\frac{1}{N_k}\sum_i
\delta(k_i,k)\frac{1}{k_i}\sum_j a_{ij}\lay{1}q_j
\end{equation}
which corresponds to the average degree at layer $2$ of the neighbors
on layer $1$ of a node having degree $k$ on layer $1$, and by the dual
expression: 
\begin{equation}
\widetilde{k}_{nn}(q)=\frac{1}{N_q}\sum_i
\delta(q_i,q)\frac{1}{q_i}\sum_j a_{ij}\lay{2}k_j
\end{equation}
Here $N_k$ (resp. $N_q$) indicates the number of nodes having degree
equal to $k$ (resp. $q$) on layer $1$ (resp. on layer $2$).  In
Fig.~\ref{fig:fig7_new}(d)-(f) we show the intra- inter- and mixed
correlation patterns obtained for several values of
$\beta$. Interestingly, for different values of the parameters one
obtains different intra- and inter-layer correlation patterns, but
also assortative or disassortative mixed correlations. 

\section{Distance and interdependence}
\label{sec:distance}

Despite the main focus of the present work is on the properties of
degree distribution and inter-layer degree correlations, we have also
explored the distribution of shortest path length and the actual
organization of shortest paths in the multiplex as a function of the
two parameters $\alpha$ and $\beta$. It is important to notice that in
a multiplex network the shortests paths between any pair of nodes are
not limited to just one layer but can instead span both
layers. Therefore, aside with the classical measure of characteristic
path length:
\begin{equation}
  \avg{l} = \frac{1}{N(N-1)}\sum_{i}\sum_{j<i}d_{ij},
\end{equation}
which is just the average over all possible pairs of nodes of the
distance $d_{ij}$ between node $i$ and node $j$, we also computed the
multiplex interdependence\cite{Morris:2012, Nicosia2013}:
\begin{equation}
  \lambda =\frac{1}{N}\sum_{i}\sum_{j\neq i}
  \frac{\psi_{ij}}{\sigma_{ij}}.
\end{equation}
The quantity $\lambda\in [0,1]$ is the average ratio between the
number $\psi_{ij}$ of shortest paths between node $i$ and node $j$
which use edges lying on both layers and the total number
$\sigma_{ij}$ of shortest paths between $i$ and $j$ in the
multiplex. When $\lambda\simeq 0$ then almost all shortest paths run
in just one layer, while at the other extreme $\lambda\simeq 1$ all
shortest paths use edges in both layers. 

In Fig.~\ref{fig:fig8_new}(a)-(b) we report the value of $\lambda$ and
$\avg{l}$ for a synthetic multiplex of $N=2000$ nodes as a function of
$\alpha$ and $\beta$. Notice that the behaviour of $\avg{l}$ closely
mirrors that of the participation ratio reported in
Fig.~\ref{fig:fig4_new}. As expected, the characteristic path length
is smaller in the condensed phase, due to the presence of condensed
nodes which are connected to virtually all the other nodes, and is
larger in the non-condensed phase. The behavior of $\lambda$ is more
interesting.  In fact, the non-condesed phase is characterised by a
relatively high interdependence, and its value (which is almost always
confined in the interval $[0.7:0.9]$) does not heavily depend on the
actual value of $\alpha$ and $\beta$. In the condensed phase, instead,
we spot two different sub-regions. For $\beta>0$ we have
$\lambda\simeq 0.5$, which is expected since in this regime the same
node is the condensed one on both layers, and half of the shortest
paths can indeed run on both layers. When $\beta<0$ the condensed
nodes on the two layers are distinct, so that all the shortest path
run on just one layer, i.e. through the condensed node of that layer,
and consequently $\lambda\simeq 0$.

\begin{figure}[!htbp]
  \begin{center}
    \includegraphics[width=3in]{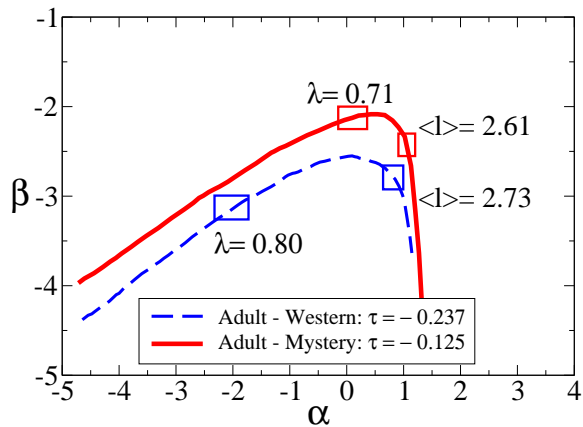}
  \end{center}
  \caption{(color online) By tuning the two exponents $\alpha$ and
    $\beta$ one can construct a synthetic multiplex network which
    reproduces some of the structural properties of real-world system.
    The solid red line and the dashed blue line indicate,
    respectively, all the $(\alpha, \beta)$ pairs which produce
    multiplex networks whose inter-layer degree correlation
    coefficient is compatible with that observed in the Adult-Mystery
    ($\tau=-0.125$) and in the Adult-Western ($\tau=-0.237$) multiplex
    networks constructed from the IMDB data set (see
    Ref.~\cite{Nicosia2014_corr}). The four boxes highlight the
    regions of the $\alpha$-$\beta$ plane in which either the
    characteristic path length $\avg{l}$ or the multiplex
    interdependence $\lambda$ are also similar to those measured on
    the Adult-Mystery and on the Adult-Western multiplexes.}
  \label{fig:fig9_new}
\end{figure}

\section{Model calibration}
\label{sec:calibration}

Here we discuss the possibility of calibrating the non-linear
preferential attachment model defined in Eq.~(\ref{f}), i.e. of
choosing appropriate values of $\alpha$ and $\beta$, in order to
reproduce some of the structural properties of a real-world multiplex
network. As an example, we consider two 2-layer multiplex networks
constructed from the IMDB costarring multi-layer network data set
described in Ref.~\cite{Nicosia2014_corr}. In this data set, each
layer corresponds to a different movie genre. The first multiplex
consists of the nodes (actors) who have acted both in Adult and in
Western movies, while the second one includes actors who have starred
both in Adult and Mystery movies. Both systems are characterized by
negative inter-layer degree correlations ($\tau=-0.237$ and
$\tau=-0.125$, respectively). Let us now imagine that we want to set
the values of $\alpha$ and $\beta$ in order to construct a synthetic
network having the same inter-layer degree correlation pattern of each
of the two multiplexes.

In Fig.~\ref{fig:fig9_new} we report the curves in the
$\alpha$-$\beta$ plane corresponding to the values of $\tau$ measured
in the Adult-Western and in the Adult-Mystery multiplexes. All the
points of the solid red curve are pairs of values $(\alpha, \beta)$
which produce a 2-layer multiplex with $\tau=-0.125$ (Adult-Mystery),
while the pairs $(\alpha,\beta)$ indicated by the dashed blue line
correspond to $\tau=-0.237$ (Adult-Western). We notice that each of
these pairs of parameters produces a synthetic multiplex network
having exaclty the same value of $\tau$, but in general different
values of characteristic path length $\avg{l}$ and interdependence
$\lambda$. In the same plot we show, for each network, the range of
values which guarantee, respectively, a value of $\avg{l}$ or a value
of $\lambda$ compatible with those observed in the real
multiplexes. This example shows that, despite being elegant and
analytically solvable, the non-linear attachment model cannot
reproduce, at the same time, all the structural properties of
real-world networks. This fact suggests that the preferential
attachment mechanism is just one among several ingredients responsible
for the formation of multiplex networks.

\section{General models for $M$ layers}
\label{sec:general}

The model defined in Eq.~(\ref{pi}) and Eq.~(\ref{f}) can be
generalized to the case of multiplexes with $M$ layers in at least
three different ways. We review them in the following, and for the
first two generalization we also give a sketch of the the derivation
of the conditions for condensation.

\subsection{One vs. All}
A simple extension would be to consider an attaching function
\begin{equation}
  f^{a}(\vec{\bm{k}}) = (k\lay{a})^{\alpha} \prod_{b\neq a}
  (k\lay{b})^{\beta}
  \label{Ga}
\end{equation}
which says that the probability for a new node to attach on layer $a$
to a node having degree $k\lay{a}$ depends on the $\alpha-$ power of
$k\lay{a}$ and on the product of the $\beta-$powers of the degrees of
the same node at the other layers $b\neq a$. In this case, each layer
can either compete with all the others ($\beta<0$) or cooperate with
all of them ($\beta>0$), and the behavior of any two layers will be
exactly the same of that studied in the previous Sections.

Following a similar approach used to determine the condensation phase
diagram for the model of two layers, it is easy to show that the
condensation occurs in a multiplex of $M$ layers satisfying the
attachment rule given by Eq.~(\ref{Ga}) under the following
conditions
\begin{equation}
\alpha>1,\  \beta<0~~~~~~ \mbox{or}~~~~~~~~\alpha+\beta (M-1)>1, \  \beta>0.
\end{equation}
In the case $\beta<0$ and $\alpha>1$ there are $M$ nodes in which the
condensation occurs, exactly one in each of the $M$ layers. Each of
these condensed nodes has degree $k^{[c]}\simeq t$ in exactly one
layer (say layer $c$), while its degree on all the other layers is
equal to $m$. Instead for $\beta>0$ and $\alpha+\beta (M-1)>1$ the
condensation occurs on a single node that has degree $k\lay{c}\simeq
t$ in all the layers of the multiplex.

\subsection{Two groups of layers}

Another possible extension of Eq.~(\ref{f}) to the case of $M$ layers
considers layers divided into two groups, say $\Gamma_1$ and
$\Gamma_2$ . We denote by $\Gamma(a)$ the group of layers to which
layer $a$ belongs, and by $M_1,M_2$ their cardinality
$M_1=|\Gamma_{1}|$ and $M_2=|\Gamma_2|$.  We define the attaching
function:
\begin{equation}
  f\lay{a}(\vec{\bm{k}}) = \prod_{b:
    \Gamma(b)=\Gamma(a)}\!\!\!\!\!(k\lay{b})^{\alpha} \prod_{b: \Gamma(b)\neq
    \Gamma(a)} \!\!\!\!\!(k\lay{b})^{\beta}
    \label{Gb}
\end{equation}
meaning that the probability for a new node to connect on layer $a$
with a node of degree $k\lay{a}$ depends on the product of the
$\alpha$-power of the degrees of the destination node at all layers
belonging to the same group of layer $a$ multiplied by the product of
the $\beta-$powers of the degrees of the destination node at all
layers belonging to the other group. Also in this case the dynamics of
pairwise relationships between layers belonging to different groups is
similar to that observed in the 2-layer case discussed in the previous
Sections. Though, the phase diagram is not exactly the same. In fact,
the condensation could occur either only on the layers belonging to
$\Gamma_1$, or only on the layers belonging to $\Gamma_2$ or on all
the $M$ layers at the same time.

Following a similar approach used to determine the condensation phase
diagram for the model of two layers, it is possible to show that the
condensation occurs in a multiplex of $M$ layers satisfying the
attachment rule given by Eq.~(\ref{Gb}) under the following conditions
\begin{eqnarray}
  &\alpha M_1>1,\   \beta<0\quad\mbox{or}\nonumber\\
  &\alpha M_2>1, \  \beta <0\quad\mbox{or}\nonumber\\
  &\xi=\alpha (M_1^2+M_2^2)+2 \beta M_1M_2>2, \  \beta>0.\quad
\end{eqnarray}
In the case $\beta<0$ and $\alpha M_1>1$ there is a node in which the
condensation occurs. This node has all the degrees in layers $c\in
\Gamma_{1}$ given by $k^{[c]}\simeq t$. Similarly for $\beta<0$ one
node becomes the condensate in layers $c\in\Gamma_{2}$ if $\alpha
M_2>1$.  If both $\alpha M_1>1$ and $\alpha M_2>1$ these two nodes
where the condensation occurs coexist in the multiplex and are
distinct.  Instead for $\beta>0$ and $\xi>2$ the condensation occurs
on a single node that have all the degrees $k^{[a]}\simeq t$ in every
layer $a$.

\subsection{More complex layer interconnections}

Finally, we consider the case in which the degree of a node at each
single layer might interact with the degree of the same node at any
other layer by means of a power $\alpha$ or $\beta$. We define a
$M\times M$ interaction matrix $\mathcal{C} = \{c_{a,b}\}$, such that
$c_{a,b}= +1$ if layer $a$ interacts with layer $b$ through the
exponent $\alpha$, while $c_{a,b}= -1$ if $a$ interacts with $b$
through the exponent $\beta$. Notice that in general $c_{a,b}\neq
c_{b,a}$, i.e. $\mathcal{C}$ is not necessarily symmetric. In this case
the attaching function reads:
\begin{equation}
  f^{a}(\vec{\bm{k}}) = \prod_{b: c_{a,b}=1}(k\lay{b})^{\alpha}
  \prod_{b: c_{a,b}=-1} (k\lay{b})^{\beta}
  \label{Gc}
\end{equation}
This model is very general and allows a pretty rich interplay between
the degree distributions of the $M$ layers. In this case the
conditions for condensation depend on the structure of the
interconnection matrix $\mathcal{C}$, and the derivation is left as a
future work.

\section{Conclusions}
\label{sec:conclusions}

In this Article we have introduced a general class of non-linear
models to grow multiplexes which display a rich variety of behaviors,
including the appearance of positive, null and negative inter-layer
degree correlations and the transition to a condensed phase. We have
shown that the model is highly sensitive to stochasticity, so that the
mean-field approach, which has been fundamental to study growth
processes on single-layer networks, fails to give account for some of
its most interesting properties. Conversely, the solution of the
master equation of the system gives some general theoretical insights
which will certainly prove to be a useful guide in the exploration of
real-world multiplex networks.

We would like to stress the fact that the class of growth models
proposed in this work includes only some of the ingredients which
might be responsible for the formation and evolution of multi-layer
networks. As a matter of fact, real networked systems rarely evolve
only by the addition of new nodes and edges at discrete
time-steps. Depending on the structure and function of the multiplex
system under study, nodes can also disappear and re-join the network
again, with a different number of edges on each layer, and edges might
be rewired, severed and re-created, sometimes according to the state
of some dynamical processes occurring on the network. Also, the
arrival and departure of nodes and the creation and rewiring of edges
might be affected by different levels of topological and temporal
correlations.  All these ingredients should be taken into account for
a more accurate modelling of real-world multiplex networks, and this
will certainly be the subject of future research in this novel field
of network science. Nevertheless we believe that, despite the few
simplifying assumptions introduced to make the model analytically
tractable, the present work clearly points out that multiplex networks
are indeed characterized by new, additional and somehow unexpected
levels of complexity, and that the multiplex perspective might reveal
interesting aspects of real-world complex systems which have remained
unnoticed until now.

\begin{acknowledgments}
  V.N. and V.L. acknowledge support from the Project LASAGNE, Contract
  No.318132 (STREP), funded by the European Commission. M.B. is
  supported by the FET-Proactive project PLEXMATH (FP7-ICT-2011-8;
  grant number 317614) funded by the European Commission. This
  research utilised Queen Mary's MidPlus computational facilities,
  supported by QMUL Research-IT and funded by EPSRC grant
  EP/K000128/1.
\end{acknowledgments}

\appendix

\renewcommand\theequation{{A-\arabic{equation}}}
\renewcommand\thetable{{A-\Roman{table}}}
\renewcommand\thefigure{{A-\arabic{figure}}}
\setcounter{figure}{0}

\section{Inter-layer correlations}

\subsection{Coefficients to quantify inter-layer degree correlations}

To detect and quantify the presence of inter--layer degree
correlations we have evaluated the Pearson's linear correlation
coefficient $r$, the Spearman rank correlation coefficient $\rho$ and
the Kendall's $\tau$ rank correlation coefficient of the degree
distributions at the two layers.
If we denote as $k_i$ and $q_i$ the degrees of node $i$ respectively
at layer $1$ and layer $2$, the Pearson's correlation coefficient of
the two degree sequences is defined as:
\begin{equation}
  r = \frac{  \avg{k q} - \avg{k} \avg{q}} {\sigma_k \sigma_q}
\end{equation}
where the averages are taken over all the nodes in each layer, and the
$\sigma_{\bullet}$ are the corresponding standard deviations.
Similarly, if we denote by $r(k_i)$ the rank of the degree of node $i$
on the first layer, and by $r(q_i)$ the rank of the degree of node $i$
on the second layer, the Spearman's correlation coefficient is defined
as:
\begin{equation}
  \rho = \frac{\sum_{i}\left(r(k_i) -
    \overline{r(k)}\right)\left(r(q_i) -
    \overline{r(q)}\right)}{\sqrt{\sum_i\left(r(k_i) -
      \overline{r(k)}\right)^2\sum_i \left(r(q_i) -
      \overline{r(q)}\right)^2}}
\end{equation}
where $\overline{r(k)}$ and $\overline{r(q)}$ are the averages respectively 
at layer $1$ and layer $2$. 

If we consider node $i=(i\lay{1}, i\lay{2})$ and $j=(j\lay{1},
j\lay{2})$ and we call $r(\cdot)$ the ranking induced at each layer by
the degree sequence, we say that $(i,j)$ is a concordant pair with
respect to $r(\cdot)$ if the ranks of the two nodes agree, i.e. if
both $r(i\lay{1}) > r(j\lay{1})$ and $r(i\lay{2}) > r(j\lay{2})$ or
both $r(i\lay{1}) < r(j\lay{1})$ and $r(i\lay{2}) < r(j\lay{2})$. If a
pair of nodes is not concordant, then it is said discordant. The
Kendall's $\tau$ coefficient measures the correlation between two
rankings by looking at concordant and discordant pairs:

\begin{equation}
  \tau=\frac{n_c - n_d}{\sqrt{(n_0 -n_1)(n_0-n_2)}}
\end{equation}

\begin{figure*}[!t]
  \begin{center}
    \includegraphics[width=6in]{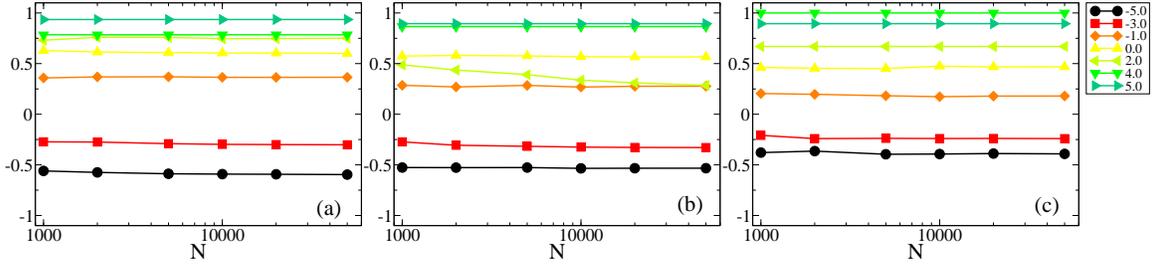}
  \end{center}
  \caption{(color online) In the three panels we show the value of the
    Kendall's $\tau$ correlation coefficient as a function of the size
    $N$ of the multiplex, respectively for (a) $alpha=-1.0$, (b)
    $\alpha=0.0$ and (c) $\alpha=1.0$, and several values of
    $\beta$. Notice that the value of $\tau$ does not depend on $N$,
    meaning that the shape of region $-_{b}$ in
    Fig.~\ref{fig:fig4_new}(c) is not due to the finite size of the
    network.}
  \label{fig:tau_sizedep}
\end{figure*}

where $n_c$ is the number of concordant pairs, $n_d$ is the number of
discordant pairs, and $n_0=1/2 N (N-1)$ is the total possible number
of pairs in a set of $N$ elements. The terms $n_1$ and $n_2$ account
for the presence of rank degeneracies. In particular, let us suppose
that the first ranking has $m$ tied groups, i.e. $m$ sets of elements
such as all the elements in one of this set have the same rank.  If we
call $u_i$ the number of nodes in the $i^{\text th}$ tied group, then
$n_1$ is defined as:
$$
n_1 = \sum_{i=1}^{m}\frac{1}{2} u_i(u_i -1).
$$ 
Similarly, $n_2$ is defined as follows:
$$
n_2 = \sum_{j=1}^{n}\frac{1}{2} v_j(v_j -1)
$$ 
where we have made the assumption that the second ranking has $n$ tied
groups, and that the $j^{\text th}$ tied group has $v_j$ elements.

The Kendall's $\tau$ coefficient is equal to $1$ when the rankings
induced by the degree sequence at each layer are perfectly concordant,
while $\tau=-1$ if one of the two rankings is exactly the opposite of
the other.

\subsection{Pearson's coefficient in the non-condensed phase}

Using the master equation we can derive several relations between the
moment of the degree distribution at long times.  In particular it can
be shown that for $m=1$ we have
\begin{eqnarray} 
  \displaystyle \Avg{k^rq^s}&=C+\Avg{(k+1)^r
  k^{\alpha}q^{s+\beta}}-\Avg{k^{r+\alpha}q^{s+\beta}}\nonumber\\
  &+\Avg{(q+1)^sq^{\alpha}
  k^{r+\beta}} -\Avg{q^{s+\alpha}k^{r+\beta}}.
\end{eqnarray} 
In particular we have, \begin{equation}
  C\Avg{k^2}=2C+2\avg{k^{1+\alpha}q^{\beta}},~~~~~~~~~~~~~
  C\Avg{kq}=2\Avg{k^{\alpha}q^{1+\beta}}+C.  \end{equation} 
Therefore the Pearson's linear coefficient $r$, defined
as
\begin{equation} 
  r=\frac{\Avg{kq}-\Avg{k}\Avg{q}}{\sigma_k \sigma_{q}} 
\end{equation}
with $\sigma_k^2=\Avg{k^2}-\Avg{k}^2$ can be also written as
\begin{equation}
  r=\frac{\Avg{k^{\alpha}q^{1+\beta}}-3/2\Avg{k^{\alpha}q^{\beta}}}{\Avg{k^{1+\alpha}q^{\beta}}-\Avg{k^{\alpha}q^{\beta}}}.
\end{equation}

\section{Stability of negative inter-layer correlations}

It is interesting to investigate whether the existence of two
sub-regions in the phase diagram of $\tau$ for $\beta<0$ (see
Fig.~\ref{fig:fig4_new}(c) in the main text) is indeed due to
finite-size effects or not. To this aim, we computed $\tau$ for
networks whose size varied across three orders of magnitude. The
results are reported in Fig.~\ref{fig:tau_sizedep}, for different
values of $\alpha$ and $\beta$. As made clear by the figures, the
values of $\tau$ measured for a certain pair $(\alpha, \beta)$ do not
depend on the size of the multiplex, and therefore the shape of the
region $-_{b}$ is not an artifact due to finite size effects.

\end{document}